\documentclass{article}
\usepackage{amsmath}
\usepackage{mathtools}
\usepackage{array}
\usepackage{url}
\usepackage{multirow}
\usepackage{tikz}
\usetikzlibrary{decorations.pathreplacing}
\newcolumntype{C}[1]{>{\centering\let\newline\\\arraybackslash\hspace{0pt}}m{#1}}

\title{Academic Torrents: Scalable Data Distribution}
\author{Henry Z. Lo\thanks{All authors contributed equally.}, Joseph Paul Cohen\footnotemark[1]}
\begin{document}
\maketitle

\section{Introduction}

As competitions get more popular, transferring ever-larger data sets becomes infeasible and costly.  For example, downloading the 157.3 GB 2012 ImageNet data set incurs about \$4.33 in bandwidth costs per download.  Downloading the full ImageNet data set takes 33 days. ImageNet has since become popular beyond the competition, and many papers and models now revolve around this data set.  For sharing such an important resource to the machine learning community, the sharers of ImageNet must shoulder a large bandwidth burden.

Academic Torrents reduces this burden for disseminating competition data, and also increases download speeds for end users \footnote{\url{http://academictorrents.com}.  Academic Torrents is run by a pending nonprofit.}.  By augmenting an existing HTTP server with a peer-to-peer swarm, requests get re-routed to get data from downloaders.  While existing systems slow down with more users, the benefits of Academic Torrents grow, with noticeable effects even when only one other person is downloading.

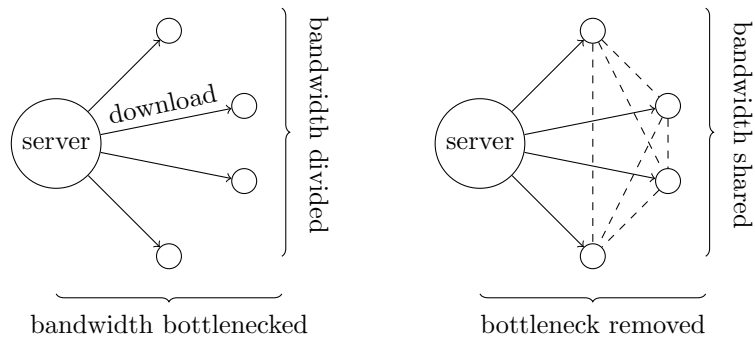
\begin{figure}
\centering
\begin{tikzpicture}
    \node[draw,circle] (root) at (0,0) {server};
    \node[draw,circle] (c1) at (1.5,1.5) {};
    \node[draw,circle] (c2) at (2.5,0.5) {};
    \node[draw,circle] (c3) at (2.5,-0.5) {};
    \node[draw,circle] (c4) at (1.5,-1.5) {};
    \path (root) edge[->] node {} (c1)
                 edge[->] node[rotate=12,above] {download} (c2)
                 edge[->] node {} (c3)
                 edge[->] node {} (c4);
    \draw[decorate,decoration={brace,amplitude=3pt,mirror}] 
    (0,-2) coordinate (0,1) -- (3,-2) coordinate (3,1); 
    \node (text1) at (1.5,-2.4) {bandwidth bottlenecked};
     \draw[decorate,decoration={brace,amplitude=3pt,mirror}] 
    (3,-1.5) coordinate (3,1) -- (3,1.8) coordinate (3,2); 
    \node [rotate=270] (text1) at (3.5,0.25) {bandwidth divided};
\end{tikzpicture}
\hspace{3em}
\begin{tikzpicture}
    \node[draw,circle] (root) at (0,0) {server};
    \node[draw,circle] (c1) at (1.5,1.5) {};
    \node[draw,circle] (c2) at (2.5,0.5) {};
    \node[draw,circle] (c3) at (2.5,-0.5) {};
    \node[draw,circle] (c4) at (1.5,-1.5) {};
    \path (root) edge[->] node {} (c1)
                 edge[->] node {} (c2)
                 edge[->] node {} (c3)
                 edge[->] node {} (c4)
          (c1)   edge[dashed] node {} (c2)
                 edge[dashed] node {} (c3)
                 edge[dashed] node {} (c4)
          (c2)   edge[dashed] node {} (c3)
                 edge[dashed] node {} (c4)
          (c3)   edge[dashed] node {} (c4);
    \draw[decorate,decoration={brace,amplitude=3pt,mirror}] 
    (0,-2) coordinate (0,1) -- (3,-2) coordinate (3,1); 
    \node (text1) at (1.5,-2.4) {bottleneck removed};
     \draw[decorate,decoration={brace,amplitude=3pt,mirror}] 
    (3,-1.5) coordinate (3,1) -- (3,1.8) coordinate (3,2); 
    \node [rotate=270] (text1) at (3.5,0.25) {bandwidth shared};
\end{tikzpicture}
\caption{HTTP (client-server) data sharing vs. HTTP + peer-to-peer.  Peer-to-peer swarms provide more data sources for downloaders, reducing server load.}
\end{figure}

\section{Case Study}

Reddit public comments\footnote{\url{https://www.reddit.com/r/datasets/comments/3bxlg7/i_have_every_publicly_available_reddit_comment/}} is a 160.68 GB data set (when compressed) of text currently linked to on Kaggle.  Since being shared on Academic Torrents in May 2015, the original seeder has uploaded 366.68 GB, while the entire community has downloaded 15.43 TB.  The upload download ratio is:
\begin{align}
\text{U/D} = \underbracket{366.68 \text{ GB}}_{\text{Sharer uploaded}} / \underbracket{15.43 \text{ TB}}_{\text{Total downloaded}} = \mathbf{42.067}
\end{align}

For every byte uploaded, the community matched that contribution 41 times.  In a standard HTTP upload / download system, sharing the same amount of data would incur the seeder 42.067 times more bandwidth.

This U/D ratio translates to real costs.  To the uploader, the Reddit data set costs \$4.42 for each download \footnote{Assuming US Amazon S3 data transfer costs of \$0.0275 per GB.}  For the 96 downloads, the associated HTTP bandwidth would mean \$424.32, while the bandwidth used on Academic Torrents reduces the bill to \$10.09.

Downloaders also get their data faster.  From our university server, we are currently downloading the entire 1.2 TB ImageNet data at 500 KB/s, resulting in a 33-day download.  On Academic Torrents, we have previously reached speeds of 34 MB/s; a download at this speed would thus finish in 9.8 hours.  This download speed is limited only by the bandwidth of the pipe.

Using this data, we cost-project the benefits of Academic Torrents with the same U/D ratio and speeds that we saw on our previous downloads (Table \ref{savings}).

\begin{table}
\centering
\resizebox{\textwidth}{!}{
\begin{tabular}{l|c|c|c|c|c|c|}
\multirow{2}{*}{\textbf{Challenge}} & \multicolumn{3}{c|}{\textbf{Upload Bandwidth (100 DLs)}} & \multicolumn{3}{c|}{\textbf{Download Speed}} \\
 & HTTP & AT & \textbf{Savings} & HTTP & AT & \textbf{Savings} \\
\hline
Whale \footnote{https://www.kaggle.com/c/noaa-right-whale-recognition/data}                   & 873.00 GB  & 20.68 GB  & \$23.36  & 4.85 h  & 0.07 m  & 4.78 h \\
Diabetes \footnote{https://www.kaggle.com/c/diabetic-retinopathy-detection/data}                & 8.22 TB  & 0.20 TB & \$220.68 & 45.66 h & 0.67  m & 44.99 h \\
ImageNet & 15.73 TB & 0.37 TB & \$422.29 & 87.39 h & 1.28 h    & 86.11 h \\
\end{tabular}}
\caption{Cost savings (for uploading) and downloading using Academic Torrents.  Values are projections based on actual measures from the Reddit public comments data set.  Upload calculations are for 100 downloads.  Time is based on 34 MB/s download speed.\label{savings}}
\end{table}

\section{Future}
Academic Torrents is maintained by the pending-nonprofit Institute for Reproducible Research.  Support for terms of service for data sets was recently added.  Academic Torrents currently indexes 10.12 TB of data, serves 30000 unique visitors a month, and facilitates over 900 GB a day.  Future efforts will scale up the platform and improve its user interface to make it more accessible.
\end{document}